\documentclass{llncs}
\usepackage{makeidx}  
\usepackage{amsfonts,amssymb}
\usepackage{graphicx}
\begin{document}

\title{A Tighter Analysis of Setcover Greedy Algorithm for Test Set}
\titlerunning{A Tighter Analysis}  
%
\author{Peng Cui\inst{1}}
\authorrunning{Cui Peng}   
%
\tocauthor{Cui Peng (Renmin University of China)}
\institute{Key Laboratory of Data Engineering and Knowledge Engineering,\\
           Renmin University of China, Beijing 100872, China,\\
\email{cuipeng@ruc.edu.cn}}

\maketitle              

\begin{abstract}
Setcover greedy algorithm is a natural approximation algorithm for test set
problem. This paper gives a precise and tighter analysis of performance
guarantee of this algorithm. The author improves the performance guarantee
$2\ln n$ which derives from set cover problem to $1.1354\ln n$ by applying the
potential function technique. In addition, the author gives a nontrivial lower
bound $1.0004609\ln n$ of performance guarantee of this algorithm. This lower
bound, together with the matching bound of information content heuristic,
confirms the fact information content heuristic is slightly better than
setcover greedy algorithm in worst case.
\end{abstract}
\section{Introduction}
The test set problem is NP-hard. The polynomial time approximation
algorithms using in practice includes "greedy" heuristics
implemented by set cover criterion or by information
criterion\cite{ms}. Test set can not be approximated within
$(1-\varepsilon)\ln n$ for any $\varepsilon>0$ unless $NP\subseteq
DTIME(n^{\log\log n})$\cite{bh,bdk}. Recently, the authors of
\cite{bdk} designed a new information type greedy algorithm,
information content heuristic (ICH for short), and proved its
performance guarantee $\ln n+1$, which almost matches the
inapproximability results.

The setcover greedy algorithm (SGA for short) is a natural
approximation algorithm for test set. In practice, its average
performance is virtually the same as information type greedy
algorithms\cite{ms,dk}. The performance guarantee $2\ln n$ of SGA is
obtained by transforming the test set problem as a set cover
problem. The authors of \cite{bh} give the tight performance
guarantee $11/8$ of SGA on instances with the size of tests no
greater than 2.

Oblivious rounding, a derandomization technique to obtain simple
greedy algorithm for set cover problems by conditional probabilities
was introduced in \cite{y}. Young observed the number of elements
uncovered is an "potential function" and the approximation algorithm
only need to drive down the potential function at each step, thus he
showed another proof of the well-known performance guarantee $\ln
n+1$.

In this paper, the author presents a tighter analysis of SGA. The author uses
the potential function technique of \cite{y} to improve the performance
guarantee $2\ln n$ which derives from set cover problem to $1.1354\ln n$, and
construct instances to give a nontrivial lower bound $1.0004609\ln n$ of the
performance guarantee. The latter result confirms the fact ICH is slightly
better than SGA in worst case. In this analysis, the author refers to the tight
analysis of the greedy algorithm for set cover problem in \cite{s}.

In Section 2, the author shows the two main theorems, and some definitions,
notations and facts are given. In Section 3, the author analyzes
differentiation distribution of item pairs and uses the potential function
method to prove the improved performance guarantee. In Section 4, the author
shows the nontrivial lower bound by constructing instances of test set with
arbitrary large size. Section 5 is some discussions.

\section{Overview}
The input of test set problem consists of $S$, a set of items (called
universe), and $\mathcal T$, a collection of subsets (called tests) of $S$. An
item pair $\{i,j\}$ is a subset of $S$ containing two different items of $S$. A
test $T$ differentiates item pair $\{i,j\}$ if $|T\cap \{i,j\}|=1$. $\mathcal
T$ is a test set of $S$, i.e. any item pair of $S$ is differentiated by one
test in $\mathcal T$. The objective is to find $\mathcal T'\subseteq \mathcal
T$ with minimum cardinality which is also a test set of $S$. We use $\mathcal
T^*$ to represent the optimal test set. Denote $n=|S|$, and $m^*=|\mathcal
T^*|$. In this paper, we assume $m^*\ge 2$.

Among an instance of test set problem, there are $n\choose 2$ different item
pairs. Let $i,j$ be two different items, and $S_{1},S_{2}$ be two disjoint
subsets of $S$. If $i,j\in S_{1}$, we say $\{i,j\}$ is an item pair inside
$S_{1}$, and if $i\in S_{1}$ and $i\in S_{2}$, we say $\{i,j\}$ is an item pair
between $S_{1}$ and $S_{2}$.

Let $S'$ be a subset of $S$, $\mathcal T$ be a collection of tests of $S$, we
say $\mathcal T$ is a test set of $S'$ iff any item pair inside $S'$ is
differentiated by one test in $\mathcal T$. Notice tests in $\mathcal T$ may
contains items that are not in $S'$. Clearly, if $\mathcal T$ is a test set of
$S$, $\mathcal T$ is a test set of $S'$.

We use $\{i,j\}\perp T$ to represent that $T$ differentiates $\{i,j\}$ and
$\{i,j\}\parallel T$ to represent that $T$ does not differentiate $\{i,j\}$. We
use $\{i,j\}\perp \mathcal T$ to represent that at least one test in $\mathcal
T$ differentiates $\{i,j\}$, $\{i,j\}\parallel \mathcal T$ to represent that
any test in $\mathcal T$ does not differentiate $\{i,j\}$, and $\perp
(\{i,j\},\mathcal T)$ to represent the number of tests in $\mathcal T$ that
differentiate $\{i,j\}$.

{\bf Fact 1.} \it For three different items $i$, $j$ and $k$, if
$\{i,j\}\parallel \mathcal T$ and $\{i,k\}\parallel \mathcal T$,
then $\{j,k\}\parallel \mathcal T$.\rm

{\bf Fact 2.} \it For three different items $i$, $j$ and $k$, and a test $T$,
if $\{i,j\}\perp T$ and $\{i,k\}\perp T$, then $\{j,k\}\parallel T$.\rm

Given $\mathcal T'\subseteq \mathcal T$, we define a binary relation
$\backsim_{\mathcal T'}$ on $S$: for two item $i,j$,
$i\backsim_{\mathcal T'}j$ iff $\{i,j\}\parallel \mathcal T'$. By
Fact 1, $\backsim_{\mathcal T'}$ is an equivalent relation. The
equivalent classes containing $i$ is denoted as $[i]$.

{\bf Fact 3.} \it If $\mathcal T$ is a minimal test set, then
$|\mathcal T|\le n-1$.\rm

{\bf Fact 4.} \it If $\mathcal T$ is a test set, then $|\mathcal
T|\ge \log_{2} n$.\rm

Test set $\mathcal T$ with $|\mathcal T|=\log_{2} n$ is
called a compact test set. If $\mathcal T$ is a compact test set,
then $|S|=2^q, q\in Z^+$.

In set cover problem, we are given $U$, the universe, and $\mathcal C$, a
collection of subsets of $U$. $\mathcal C$ is a set cover of $U$, i.e.
$\bigcup_{c\in \mathcal C}=U$. The objective is to find $\mathcal C'\subseteq
\mathcal C$ with minimum cardinality which is also a set cover of $S$.

The greedy algorithm for set cover runs like that. In each
iteration, simply select a subset covering most uncovered elements,
repeat until all elements are covered, and return the set of
selected subsets. Let $N$ be the size of the universe, and $M^*$ be
the size of the optimal set cover. The greedy algorithm for set
cover has performance guarantee $\ln N-\ln\ln N+\Theta(1)$ by
\cite{s}.

We give two lemmas about the greedy algorithm for set cover. Lemma 1
is a corollary of Lemma 2 in \cite{s} and Lemma 2 is a corollary of
Lemma 1 and Lemma 4 in \cite{s}.

{\bf Lemma 1.} \it The size of set cover
returned by the greedy algorithm is at most $M^*(\ln N-\ln M^*+1)$.\rm

{\bf Lemma 2.} \it Given $N$ and $M^*$, there are instance of set
cover problem such that the size of set cover returned by the greedy
algorithm is at least $(M^*-1)(\ln N-\ln M^*)$.\rm

Test set problem can be transformed to set cover problem in a
natural way. Let $(S,\mathcal T)$ be an instance of test set, we
construct an instance $(U,\mathcal C)$ of set cover, where
$U=\{\{i,j\}|i,j\in S,i\ne j\}$, and $\mathcal C=\{c(T)|T\in
\mathcal T\},c(T)=\{\{i,j\}|i\in T,j\in S-T\}.$

Clearly,  $\mathcal T'$ is a test set of $S$ iff $\mathcal
C'=\{c(T)|T\in \mathcal T'\}$ is a set cover of $U$.

SGA can be described as:\\

\indent{\bf Input:} $S$,$\mathcal T$;\\
\indent{\bf Output:} a test set of $S$;\\
\indent$\bar\mathcal T\leftarrow\varnothing$;\\
\indent {\bf while} $\#(\bar\mathcal T)>0$ {\bf do}\\
\indent\indent select $T$ in $\mathcal T-\bar\mathcal T$
minimizing $\#(\bar\mathcal T\cup\{T\})$;\\
\indent\indent $\bar\mathcal T\leftarrow\bar\mathcal T\cup\{T\}$;\\
\indent {\bf endwhile}\\
\indent {\bf return} $\bar\mathcal T$;\\
\rm

In SGA, we call $\bar\mathcal T$ the partial test set. The
differentiation measure of $\bar\mathcal T$, $\#(\bar\mathcal T)$,
is defined as the number of item pairs not differentiated by
$\bar\mathcal T$. The differentiation measure of $T$ w.r.t.
$\bar\mathcal T$ is defined as $\#(T,\bar\mathcal T)=\#(\bar\mathcal
T)-\#(\bar\mathcal T\cup \{T\})$.

SGA is isomorphic to the greedy algorithm for set cover under the
natural transformation. Thus we immediately obtain the performance
guarantee $2\ln n$ of SGA. This paper shows a better performance
guarantee and a nontrivial lower bound of performance guarantee. The
two main theorems are:

{\bf Theorem 1.} \it The performance guarantee of SGA can be
$1.1354\ln n$. \rm

{\bf Theorem 2.} \it There are arbitrarily large instances of test set problem
such that the performance ratio of SGA on these instances is at least
$1.0004609\ln n$. \rm

In this paper, denote $[n]:=\{1,2,\cdots,n\}$. Denote $\phi(x):=\frac{1}{x}(\ln
x-1)$. The harmonious number is defined as $H_n:=\sum_{i=1}^n{\frac{1}{i}}$.

Two inequalities are listed here for convenience of proof in Section 3.

{\bf Fact 5.} \it For any $0<x<1$, $(1-x)^{1/x}<1/e$. \rm

{\bf Fact 6.} \it For any $x>1$, $\phi(x)\le 1/e^2=0.135\cdots$. \rm

\section{Improved Performance Guarantee}

\subsection{Differentiation Distribution}
In this subsection, the author analyzes the distribution of times for
which item pairs are differentiated in instances of test set,
especially the relationship between the differentiation distribution
and the size of the optimal test set.

{\bf Lemma 3.} \it Given two disjoint subsets $S_1,S_2\subseteq S$, and
$\mathcal T$, a set of tests of $S$, suppose $\mathcal T$ is a test set of
$S_1$ and a test set of $S_2$ , then at most $\min(|S_1|,|S_2|)$ item pairs
between $S_1$ and $S_2$ are not
differentiated by any test in $\mathcal T$.\\
\indent Proof. \rm Suppose $|S_1|\le |S_2|$. We claim for any item $i\in S_1$,
there is at most one item $j$ in $S_2$ satisfying $\{i,j\}\parallel \mathcal
T$. Otherwise there are two different items $j,k$ in $S_2$ such that
$\{i,j\}\parallel \mathcal T$ and $\{i,k\}\parallel \mathcal T$, then by Fact 1
, $\{j,k\}\parallel \mathcal T$, which contradicts $\mathcal T$ is a test set
of $S_2$. $\Box$

{\bf Lemma 4.} \it At most $n\log_{2}n$ item pairs are
differentiated by exactly one test in $\mathcal T^*$.\\
\indent Proof. \rm Let $B$ be the set of item pairs that are
differentiated by exactly one test in $\mathcal T^*$. We prove
$|B|\le n\log_{2}n$ by induction. When $n=1$, $|B|=n\log_{2}n$.
Suppose the lemma holds for any $n\le h-1$, we prove the lemma holds
for $n=h$.

Select $T\in \mathcal T^*$ such that $T\neq \varnothing$ and $T\neq
S$, then $|T|\le h-1$, $|S-T|\le h-1$. Since $\mathcal T^*$ is a
test set of $T$, by induction hypothesis, at most $|T|\log_{2}|T|$
item pairs inside $T$ are differentiated by exactly one test in
$\mathcal T^*$. Similarly, at most $|S-T|\log_{2}|S-T|$ item pairs
inside $S-T$ are differentiated by exactly one test in $\mathcal
T^*$.

By Lemma 3, at most $\min(|T|,|S-T|)$ item pairs between $T$ and
$S-T$ are not differentiated by any test in $\mathcal T^*-\{T\}$.
Therefore at most $\min(|T|,|S-T|)$ item pairs between $T$ and
$S-T$ are differentiated by exactly one test in $\mathcal T^*$.

W.l.o.g, suppose $|T|\le |S-T|$, then
\begin{eqnarray*}
&|B|&\le|T|\log_{2}|T|+|S-T|\log_{2}|S-T|+|T|\\
&&=|T|\log_{2}(2|T|)+|S-T|\log_{2}|S-T|\\
&&\le|T|\log_{2}|S|+|S-T|\log_{2}|S|\\
&&=|S|\log_{2}|S|.
\end{eqnarray*}
$\Box$

{\bf Lemma 5.} \it Given $S''\subseteq S'\subseteq S$, and $\mathcal T$, a set
of tests of $S'$, suppose $\mathcal T$ is a test set of $S''$ and a test set of
$S'-S''$ , then at most $|S'|\log_2{|S'|}$ item pairs between $S''$ and
$S'-S''$ are differentiated by exactly one test in $\mathcal T$.\\

\indent Proof. \rm Let $B$ be the set of item pairs between $S''$ and $S'-S''$
which are differentiated by exactly one test in $\mathcal T$. We prove that
$|B|\le |S'|\log_2{|S|'}$ by induction. When $|S|=1$ and $|S|=2$, the lemma
holds. Suppose the lemma holds for any $|S|\le h-1$, $h\ge 3$, we prove the
lemma holds for $|S|=h$.

Select $T\in \mathcal T$ such that $T\neq \varnothing$ and $T\neq S'$,
 then $|T|\le h-1$, $|S'-T|\le h-1$ (see Figure 1).
Since $\mathcal T-\{T\}$ is a test set of $S''\cap T$ and a test
set of $(S'-S'')\cap T$, by induction hypothesis, at most
$|T|\log_2{|T|}$ item pairs between $S''\cap T$ and $(S'-S'')\cap T$
are differentiated by exactly one test in $\mathcal T$. Similarly,
at most $|S'-T|\log_{2}|S'-T|$ item pairs between $S''\cap (S'-T)$
and $(S'-S'')\cap (S'-T)$
 are differentiated by exactly one test in $\mathcal T$.

\begin{figure}
\begin{center}
\includegraphics[width=0.5\textwidth,bb=140 290 440 560]{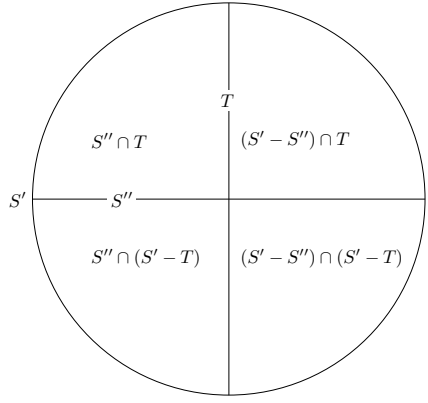}
\caption{illustration of Lemma 5}
\end{center}
\end{figure}

Since $\mathcal T-\{T\}$ is a test set of $S''\cap T$ and a test set
of $(S'-S'')\cap (S'-T)$, by Lemma 3, at most $\min(|S''\cap
T|,|(S'-S'')\cap (S'-T)|)$ item pairs between $S''\cap T$ and
$(S'-S'')\cap (S'-T)$ are not differentiated by any test in
$\mathcal T-\{T\}$. Hence at most $\min(|S''\cap T|,|(S'-S'')\cap
(S'-T)|)$ item pairs between $S''\cap T$ and $(S'-S'')\cap (S'-T)$
are differentiated by exactly one test in $\mathcal T$. Similarly,
at most $\min(|(S'-S'')\cap T|,|S''\cap (S'-T)|)$ item pairs between
$(S'-S'')\cap T$ and $S''\cap (S'-T)$ are differentiated by exactly
one test in $\mathcal T$

Clearly,
\begin{eqnarray*}
&|T|\ge&\min(|S''\cap T|,|(S'-S'')\cap (S'-T)|)\\
&&+\min(|(S'-S'')\cap T|,|S''\cap (S'-T)|).
\end{eqnarray*}

W.l.o.g, suppose $|T|\le |S'-T|$, then
\begin{eqnarray*}
&|B|&\le|T|\log_{2}|T|+|S'-T|\log_{2}|S'-T|+|T|\\
&&=|T|\log_{2}(2|T|)+|S'-T|\log_{2}|S'-T|\\
&&\le|T|\log_{2}|S'|+|S'-T|\log_{2}|S'|\\
&&=|S'|\log_{2}|S'|.
\end{eqnarray*}
$\Box$

{\bf Lemma 6.} \it At most $n\log_2{n}{m^*}^{t-1}$ item pairs are
differentiated by exactly $t$ test in $\mathcal T^*$, where $t\ge
2$.\\
\indent Proof. \rm Let $B_{t}$ be the set of item pairs that are
differentiated by exactly $t$ test in $\mathcal T^*$. For any
combination $\pi$ of $t-1$ tests in $\mathcal T^*$, let $B_{\pi}$
be the subset of $B_{t}$ such that each item pair in $B_{\pi}$ is
differentiated by any test in $\pi$.

Let $\backsim_{\pi}$ be the equivalent relation induced by
 $\pi$. For any equivalent class $[i]$, there is exactly one
 equivalent class $[j]$, such that each item pair between $[i]$
 and $[j]$ is differentiated by any test in $\pi$ (Fact 2).

Since $\mathcal T^*-\pi$ is a test set of $[i]$ and a test set of
$[j]$, by Lemma 5, at most $(|[i]\cup[j]|)\log_2{|[i]\cup[j]|}$ item
pairs between $[i]$ and $[j]$ are differentiated by exactly one test
in $\mathcal T^*-\pi$. In another word, at most
$(|[i]\cup[j]|)\log_2{|[i]\cup[j]|}$ item pairs between $[i]$ and
$[j]$ are differentiated by exactly $t$ tests in $\mathcal T^*$.
Hence
\begin{eqnarray*}
&|B_{\pi}|&\le\sum_{[i],[j]}{|[i]\cup[j]|\log_2{|[i]\cup[j]|}}\le n\log_2{n}.
\end{eqnarray*}

Therefore,
\begin{eqnarray*}
&|B_{t}|&\le\sum_{\pi}{|B_{\pi}|}\le{m^*\choose
{t-1}}n\log_2{n}\le n\log_2{n}{m^*}^{t-1}.
\end{eqnarray*}
$\Box$

{\bf Lemma 7.} \it At most $2n\log_2{n}{m^*}^{t-1}$ item pairs
are differentiated by at most $t$ test in $\mathcal T^*$, where
$t\ge 2$.\\
\indent Proof. \rm Let $B$ be the set of item pairs that are
differentiated by at most $t$ test in $\mathcal T^*$, and $B_{t}$ be
the set of item pairs that are differentiated by exactly $t$ test in
$\mathcal T^*$. By Lemma 6,
\begin{eqnarray*}
&|B|&=|B_{1}|+|B_{2}|+\cdots+|B_{t}|\\
&&\le n\log_2{n}(1+m^*+\cdots+{m^*}^{t-1})\\
&&\le 2n\log_2{n}{m^*}^{t-1}.
\end{eqnarray*}
gu $\Box$

\subsection{Proof of Theorem 1}
In this subsection, the author uses the potential function technique
to derive improved performance guarantee of SGA for test set. Our
proof is based on the trick to "balance" the potential function by
appending a negative term to the differentiation measure.

Let $I=\lceil \ln\frac{n-1}{4\log_2{n}}/\ln m^*\rceil$, then
$2n\log_2{n}{m^*}^{I-1}<{n\choose 2}\le 2n\log_2{n}{m^*}^I$. Let
$\#_{0}=1$, $\#_{1}=n\log_2{n}$, $\#_{t}=2n\log_2{n}{m^*}^{t-1},2\le
t\le I$, and $\#_{I+1}=n(n-1)/2$. Let $k_{t}=\frac{m^*}{t}\ln
\frac{t\#_{t}}{\#_{t-1}}$, $2\le t\le I+1$.

Denote by $p$ the probability distribution on tests in $\mathcal
T^*$ drawing one test uniformly from $\mathcal T^*$. For any $T\in
\mathcal T^*$ , the probability of drawing $T$ is
$p(T)=\frac{1}{m^*}$.

We divide a run of the algorithm into $I+1$ phases. For $I+1\ge t\ge
1$, Phase $t$ begins when $\#(\bar\mathcal T)\ge\#_{t-1}$ and lasts
until $\#(\bar\mathcal T)<\#_{t-1}$. Phase $t$ is blank if when
Phase $t+1$ ends, $\#(\bar\mathcal T)<\#_{t-1}$.

Let the set of selected tests in Phase $t$ is $\mathcal T_{t}$, the
partial test set when Phase $t$ ends is $\bar\mathcal T_{t}$, and
the returned test set is $\mathcal T'$. Then $\bar\mathcal
T_{t}=\cup_{t\le s\le I+1}\mathcal T_{s}$, $1\le t\le I+1$, and
$\mathcal T'=\bar\mathcal T_{2}\cup\mathcal T_{1}$. Set
$\bar\mathcal T_{I+2}=\varnothing$. If Phase $t$ is not blank, let
the last selected test in Phase $t$ is $T'_{t}$.

In Phase $t$, $I+1\ge t\ge 2$, define the potential function as
$$f(\bar\mathcal T)=(\#(\bar\mathcal T)-\textstyle\frac{t-1}{t}\displaystyle\#_{t-1})(1-
\frac{t}{m^*})^{k_{t}-|\bar\mathcal T-\bar\mathcal T_{t+1}|}.$$

By the definition of $\bar\mathcal T_{t+1}$ and Fact 5,
$$f(\bar\mathcal T_{t+1})<(\#_{t}-\textstyle\frac{t-1}{t}\displaystyle\#_{t-1})(1-\frac{t}{m^*})^{k_{t}}
<\frac{\#_{t-1}}{t}.$$

By the definition of $f(\bar\mathcal T)$ and the facts $p(T)\ge 0$
and $\sum_{T\in \mathcal T^*}{p(T)}=1$,
\begin{eqnarray*}
&&\min_{T\in \mathcal T}{f(\bar\mathcal T\cup \{T\})}\\
&&\le\min_{T\in \mathcal T^*}{f(\bar\mathcal T\cup \{T\})}\\
&&\le\sum_{T\in \mathcal T^*}({p(T)}f(\bar\mathcal T\cup
\{T\}))\\
&&=(\#(\bar\mathcal
T)-\textstyle\frac{t-1}{t}\displaystyle\#_{t-1}-\sum_{T\in \mathcal
T^*}{(p(T)\#(T,\bar\mathcal
T))})(1-\frac{t}{m^*})^{k_{t}-|\bar\mathcal T-\bar\mathcal
T_{t+1}|-1}
\end{eqnarray*}
and
\begin{eqnarray*}
&&\sum_{T\in \mathcal
T^*}{(p(T)\#(T,\bar\mathcal T))}\\
&&=\sum_{\{i,j\}\parallel \bar\mathcal T}{\sum_{T\in
\mathcal T^*:\{i,j\}\perp T}{p(T)}}\\
&&\ge\sum_{\{i,j\}\parallel \bar\mathcal T}{\frac{t}{m^*}}
-\sum_{\{i,j\}\parallel \bar\mathcal T:\perp(\{i,j\},\mathcal T^*)\le t-1}{\frac{t-1}{m^*}}\\
&&\ge(\#(\bar\mathcal
T)-\textstyle\frac{t-1}{t}\displaystyle\#_{t-1})\frac{t}{m^*}
\end{eqnarray*}
by Lemma 4 and Lemma 7.

Therefore,
$$\min_{T\in \mathcal T}{f(\bar\mathcal T\cup\{T\})}\le(\#(\bar\mathcal
T)-\textstyle\frac{t-1}{t}\displaystyle\#_{t-1})(1-\frac{t}{m^*})^{k_{t}-|\bar\mathcal
T-\bar\mathcal T_{t+1}|}=f(\bar\mathcal T).$$

During Phase $t$, the algorithm selects $T$ in $\mathcal T$ to
minimize $f(\bar\mathcal T\cup\{T\}$). Therefore, $f(\bar\mathcal
T_{t}-\{T'_{t}\})\le f(\bar\mathcal T_{t+1})<\frac{\#_{t-1}}{t}$.

On the other hand, $\#(\bar\mathcal T_{t}-\{T'_{t}\})\ge \#_{t-1}$
by definition of Phase $t$. Hence
$$f(\bar\mathcal T_{t}-\{T'_{t}\})=\frac{\#_{t-1}}{t}(1-\frac{t}{m^*})^{k_{t}-|\mathcal
T_{t}-\{T'_{t}\}|}.$$

Therefore, $(1-\frac{t}{m^*})^{k_{t}-|\mathcal
T_{t}-\{T'_{t}\}|}<1$, $|\mathcal T_{t}-\{T'_{t}\}|<k_{t}$, and
$|\mathcal T_{t}|<k_{t}+1$.

To sum up,
\begin{eqnarray*}
&|\bar\mathcal T_{2}|&<\sum_{2\le t\le I+1}{k_{t}}+I\\
&&=m^*(\sum_{2\le t\le I+1}{\frac{1}{t}\ln\frac
{\#_{t}}{\#_{t-1}}}+\sum_{2\le t\le I+1}{\frac{\ln t}{t}})+I.
\end{eqnarray*}

When all Phase $t$, $I+1\ge t\ge 2$, end, we obtain an instance of
set cover $(U,\mathcal C)$, where $U=\{\{i,j\}|\{i,j\}\parallel
\bar\mathcal T_{2}\}$ and $\mathcal C=\{c(T)|c(T)\cap U\ne
\varnothing\}$. Clearly, $|U|<\#_{1}$. Let $M^*$ be the size of the
optimal set cover of this instance. Then $|M^*|\le m^*$.

Consider the following two cases: (a)$|M^*|\le \frac{m^*}{2}$;
(b)$|M^*|>\frac{m^*}{2}$.

In case (a),
$$|\mathcal T_{1}|\le M^*(\ln\#_{1}+1)\le m^*(\textstyle\frac12\displaystyle+o(1))\ln n,$$

\noindent and
\begin{eqnarray*}
&|\bar\mathcal T_{2}|&\le m^*(\sum_{2\le t\le I+1}{\frac12\ln\frac
{\#_{t}}{\#_{t-1}}}+\frac12 \ln^2 (I+2))+I\\
&&=m^*(\textstyle\frac12\displaystyle+o(1))\ln n.
\end{eqnarray*}

Hence
$$|\mathcal T'|=m^*(1+o(1))\ln n.$$

In case (b), by Lemma 1,
\begin{eqnarray*}
&|\mathcal T_{1}|&\le M^*(\ln\#_{1}-\ln M^*+1)=m^*((1+o(1))\ln n-\ln
m^*),
\end{eqnarray*}

\noindent and
\begin{eqnarray*}
&|\bar\mathcal T_{2}|&< m^*(H_{I+1}\ln m^*+\frac12\ln 2+\frac12 \ln^2 (I+2))+I\\
&&\le m^*(\ln \frac{\ln n}{\ln m^*}\ln m^*+o(1)\ln n).
\end{eqnarray*}

Notice $\frac{\ln n}{\ln m^*}>1$, by Fact 3.

By Fact 6,
$$|\mathcal T'|\le m^*(1+\phi(\frac{\ln n}{\ln m^*})+o(1))\ln n
\le m^*(1.13533\cdots+o(1))\ln n.$$

\section{Lower Bound}
In this section, we discuss a variation of test set problem. Given disjoint
sets $S^{1},\cdots,S^{r}$ and $\mathcal T$, set of subsets of the universe
$S=S^{1}\cup\cdots\cup S^{r}$, we seek $\mathcal T'\subseteq \mathcal T$  with
minimum cardinality which is a test set of any $S^{p}$ for $1\le p \le r$.
Denote the instance by $(S^p;\mathcal T)$.

In our construction, $r=2N-J!2^{q}$, let $2^{k-1}<N^2\le 2^{k}$, we could use
the copy-split trick similar to that used in \cite{bh} to make $2^{k}$ copies
of $S^{1},\cdots,S^{r}$, split all copies by $k$ tests and split all $S^{p}$
for each copy by $r-1$ tests. Thus the splitting overhead for each copy could
be ignored.

Suppose $\hat\mathcal T$ is a compact test set of $[2^q]$. For example, we can
let $\hat\mathcal T=\{T_{k}|k\in[q]\}$, where $T_{k}$ contains integer $x$
between $1$ and $2^q$ such that the $k$-th bit of $x$'s binary representation
is $1$.

\subsection{Atom Instances}
Firstly, we give the level-$t$ atom instances $(S_t^{y};\mathcal T)$. The
universe $S_t$ includes integral points in $(t+1)$-dimension Euclid space.\\

{\bf Construction of Atom Instances.} $S_t=\bigcup_{y}{S_t^{y}}$.
$S_t^{y}=\{(x_{1},\cdots,x_{t},y)|x_{i}\in [2^q]\}$, $1\le y\le 2^{q-2}$.
$\mathcal T_t=\mathcal T^*_t\cup \mathcal T'_t$. $\mathcal T^*_t=\mathcal
T^*_{t,1}\cup\cdots\cup\mathcal T^*_{t,t}$. $\mathcal
T^*_{t,i}=\{T^*_{t,i;j}|j\in[2^{q}]\}$, $1\le i\le t$. $\mathcal T'_t=\mathcal
T'_{t,1}\cup\cdots\cup\mathcal T'_{t,t}$, $\mathcal
T'_{t,i}=\{T'_{t,i;j,k}|j\in[2^{q-2}],k\in[q]\}$, $1\le i\le t$.

$|\mathcal T^*_t|=t 2^{q}$, and $|\mathcal T'_{t}|=\frac{q}{4}t 2^{q}$.
$T^*_{t,i;j}$ contains points in $S_t^{y}$ with $x_{i}=j$. $T'_{t,i;j,k}$
contains points in $S_t^{j}$ with $x_{i}$ in the $k$-th test in $\hat\mathcal
T$. We assign an order to tests $T'_{t,i;j,k}$ in $\mathcal T'_t$, called {\it
natural order}, as the
lexical order of $(i,j,k)$.\\

An atom instance with $q=3$ and $t=2$ is shown in Figure 2.

\begin{figure}
\begin{center}
\includegraphics[width=1.0\textwidth,bb=70 330 480 520]{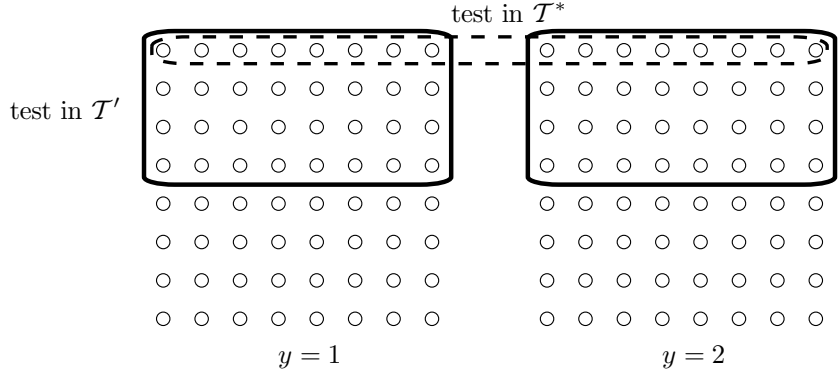}
\caption{atom instance with $q=3$ and $t=2$}
\end{center}
\end{figure}

We claim SGA could return $\mathcal T'_t$ according to their natural order in
the atom instance.

At the beginning of the algorithm, the differentiation measure of tests in
$\mathcal T'_t$ is $2^{2qt-2}$, and the differentiation measure of tests in
$\mathcal T_t^*$ is
$$2^{q(t-1)}(2^{qt}-2^{q(t-1)})2^{q-2}=2^{2qt-2}(1-2^{-q}).$$

The algorithm could first select tests in $\mathcal T'_{t,1}$ according to
their natural order. After that, the differentiation measure of tests in
$\mathcal T'_t-\mathcal T'_{t,1}$ decreases by a factor $2$, and the
differentiation measure of tests in $\mathcal T_t^*$ decreases by a factor at
least $2$. Hence, the algorithm could subsequently select tests in $\mathcal
T'_{t,2},\cdots,\mathcal T'_{t,t}$ according to their natural order.

\subsection{Level-$t$ Instances}
Secondly, we construct a series of level-$t$ instances $(S_{t}^{y,z,w};\mathcal
T_{t})$, $1\le t\le J$ based on the atom instances. Tests in the atom instances
are {\it stretched} in $w$ dimension and {\it cloned} in $z$ dimension. Let
$M^*=J!2^q$, and $N=J!2^{q(J+1)}$. The universe $S_t$ includes $N$ integral
points in $(t+3)$-dimension Euclid space.\\

{\bf Construction of Level-$t$ Instances.} $S_t=\bigcup_{y,z,w}{S_t^{y,z,w}}$.\\
$S_t^{y,z,w}=\{(x_{1},\cdots,x_{t},y,z,w)|x_{i}\in[2^q]\}$, $1\le y\le
2^{q-2},1\le z\le\frac{J!}{t},1\le w\le t2^{q(J-t)+2}$. $\mathcal T_t=\mathcal
T^*_t\cup \mathcal T'_t$. $\mathcal T^*_t=\mathcal
T^*_{t,1}\cup\cdots\cup\mathcal T^*_{t,t}$. $\mathcal
T^*_{t,i}=\{T^*_{t,i;j,l}|j\in[2^{q}],l\in[\frac{J!}{t}]\}$, $1\le i\le t$.
$\mathcal T'_t=\mathcal T'_{t,1}\cup\cdots\cup\mathcal T'_{t,t}$, $\mathcal
T'_{t,i}=\{T'_{t,i;j,k,l}|j\in[2^{q-2}],k\in[q],l\in[\frac{J!}{t}]\}$, $1\le
i\le t$.

$|\mathcal T^*_t|=M^*$, and $|\mathcal T'_{t}|=\frac{qM^*}{4}$. $T^*_{t,i;j,l}$
contains points in $S_t^{y,l,w}$ with $x_{i}=j$ for any $w$. $T'_{t,i;j,k,l}$
contains points in $S_t^{j,l,w}$ with $x_{i}$ in the $k$-th test in
$\hat\mathcal T$ for any $w$. We assign an order to tests $T'_{t,i;j,k,l}$ in
$\mathcal T'_t$,
called {\it natural order}, as the lexical order of $(i,j,k,l)$.\\

We claim SGA could select all tests in $\mathcal T'_{t,1}$ according to their
natural order in the first phase of the algorithm.

At the beginning of the algorithm, it is easy to prove that the differentiation
measure of tests in $\mathcal T'_{t,1}$ is $\#^{begin}_t=2^{q(t-1)}N$, and the
differentiation measure of tests in $\mathcal T_t^*$ is
$2^{q(t-1)}(1-2^{-q})N$.

The algorithm could select tests in $\mathcal T'_{t,1}$ according to their
natural order while the differentiation measure of the selected test is kept
equal to the differentiation measures of tests in $\mathcal T'_{t,i}$ for $2\le
i\le t$ and no less than the differentiation measure of any test in $\mathcal
T_t^*$.

When the algorithm select the last test in $\mathcal T'_{t,1}$, its
differentiation measure is $\#^{end}_t=2\cdot 2^{q(t-2)}N$.

The algorithm could subsequently select tests in $\mathcal
T'_{t,2},\cdots,\mathcal T'_{t,t}$ according to their natural order, and
returns $\mathcal T'_t$.

We explicitly show the three claims for proof of Section 4.4. \\

{\bf Claim 1.} \it In each step of SGA, the differentiation measure of the
selected test is no less than the differentiation measure of a subsequent test
according to their natural order.\rm \\

{\bf Claim 2.} \it In each step of SGA, the differentiation measure of the
selected test in every step is no less than any test in $\mathcal T_t^*$.\rm\\

{\bf Claim 3.} \it In the first phase of SGA, the differentiation measure of
the selected test in every step is at most $\#^{begin}_t$ and at least
$\#^{end}_t$.\rm

\subsection{Complete Instances}
Let $(U,\mathcal C)$ be the instance in Lemma 2, $U=\{e_{1},\cdots,e_{N}\}$,
$\mathcal C^*$ be the optimal set cover, and $\mathcal C'$ be the set cover
returned by the greedy algorithm. Construct an instance of test set
$(S_0^{p};\mathcal T_{0})$. $S_0=\bigcup_{p}{S_0^p}$.
$S_0^{p}=\{e_{p},f_{p}\}$, $1\le p\le N$. $\mathcal T_{0}=\mathcal
T'_{0}\cup\mathcal T^*_{0}$. $\mathcal T'_{0}=\mathcal C'$. $\mathcal
T^*_{0}=\mathcal C^*$.

On $(S_0^{p};\mathcal T_{0})$, the algorithm could select all the tests in
$\mathcal T'_{0}$, the differentiation measure of selected tests ranges from
$\#^{begin}_0=N/M^*$ to $\#^{end}_0=1$ by the proof of Lemma 1 in \cite{s}.

Consequently, we construct a series of level-$t$ instances
$(S_{t}^{y,z,w};\mathcal T_{t})$, $1\le t\le J$. We modify tests in $\mathcal
T'_{t,1}$ by two operations: Enlargement and Merging. In the Enlargement
operation, tests in $\mathcal T'_{t,1}$ are enlarged by a factor $2$. In the
Merging operation, tests in $\mathcal T'_{t,i}$ for $i\ge 2$ are merged to
tests in $\mathcal T'_{s,1}$ for $t-1\ge
s\ge 1$. \\

{\bf Enlargement.} Let $\mathcal T'_{t,1}=\{T'_{t,1;j,k,l}|j\in[2^{q-3}],k\in[
q],l\in[\frac{J!}{t}]\}$ for $J\ge t\ge 1$. $T_{t,1;j,k,l}$ contains points in
$S_t^{2j-1,l,w}$ and $S_t^{2j,l,w}$ with $x_{1}$ in the $k$-th test in
$\hat\mathcal T$ for any $w$. As a result, $|\mathcal
T'_{t,1}|=\frac{qM^*}{8t}$.

{\bf Merging.} By the decreasing order of $t$ for $J\ge t\ge 2$, merge tests in
$\mathcal T'_{t}-\mathcal T'_{t,1}$ by their natural order one-by-one to tests
in $\mathcal T'_{s,1}$ for $t-1\ge s\ge 1$ by the decreasing order of $s$
(primarily) and their natural order in $\mathcal T'_{s,1}$ until tests in
$\mathcal T'_{t}-\mathcal T'_{t,1}$ are exhausted.\\

The illustration of the Merging operations with $J=4$ is shown in Figure 3.

\begin{figure}
\begin{center}
\includegraphics[width=0.6\textwidth,bb=160 310 460 510]{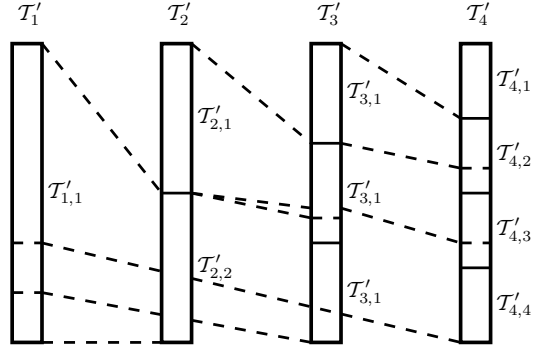}
\caption{Merging operation with $J=4$}
\end{center}
\end{figure}

For any $J\ge t\ge 2$, tests in $\mathcal T'_{s,1}$ suffice in the Merging
operation, provided that
$$|\mathcal T'_{t}-\mathcal T'_{t,1}|=\frac14(1-\frac{1}{t})qM^*
\le\frac{1}{8}H_{t-1}qM^*=|\bigcup_{s=1}^{t-1}\mathcal T'_{s,1}|.$$

Let $\mathcal T'=\mathcal T'_{0}\cup\bigcup_{t=1}^{J}\mathcal T'_{t,1}$ after
the two operations are performed. Note that after the Merging operation, tests
in $\mathcal T'_{t,1}$ for $J-1\ge t\ge 1$ will contain items from $S_{s}$ for
all $J\ge s\ge t$. Let $\mathcal T=\mathcal T^*\cup\mathcal T'$.

The complete instance is $(S_0^{p},S_{t}^{y,z,w};\mathcal T^*\cup \mathcal
T')$. Let $S=\bigcup_{t=0}^{J}{S_t}$, then $n=|S|=(J+1)N$. Suppose $\mathcal
T^*_{t}=\{T^*_{t,1},\cdots,T^*_{t,M^*}\}$, $0\le t\le J$, let $\mathcal
T^*=\{T^*_{0,1}\cup\cdots\cup T^*_{J,1},\cdots,T^*_{0,M^*}\cup\cdots\cup
T^*_{J,M^*}\}$. Since $\mathcal T^*_{0}$ is an optimal test set of $S_0$,
$\mathcal T^*$ is an optimal solution of the complete instance, and
$m^*=|\mathcal T^*|=M^*$.

\subsection{Proof of Theorem 2}
We intend to prove the performance ratio of SGA on this instance is at least
$(1+\frac{1}{J+1}(\frac{H_{J}}{8\ln 2}-1)-o(1))\ln n$, for fixed $J$. When
$J=391$, this performance ratio is at least $1.0004609\ln n$. To accomplish the
proof, we analyze the behavior of SGA on the complete instance. Indeed, the
sequence of the selected tests projected onto $S_t$ is almost the same as the
selected tests of SGA on the original level-$t$ instance for $J\ge t\ge 1$,
except that the tests in $\mathcal T'_{t,1}$ are enlarged.

Before the algorithm selects a test, let $\#_t$ be the maximum differentiation
measure of tests in $\mathcal T'_{t,1}$ for $J\ge t\ge 1$, and $\#^*$ to the
maximum differentiation measure of tests in $\mathcal T^*$.

Let $\#_{t,s}$ be the number of item pairs inside of $S_s$ contributing to
$\#_t$ for $t\le s\le J$, $\#_{t-1,s}$ be the number of item pairs inside $S_s$
contributing to $\#_{t-1}$ for $t-1\le s\le J$,  $\#_s^*$ be the number of item
pairs inside $S_s$ contributing to $\#^*$ for $1\le s\le J$, , and $\#_0^*$ be
the number of item pairs inside $S_0^{p}$ contributing to $\#^*$. Then
$\#_t=\sum_{s=t}^{J}{\#_{t,s}}$, $\#_{t-1}=\sum_{s=t-1}^{J}{\#_{t-1,s}}$, and
$\#^*=\sum_{s=0}^{J}{\#_s^*}$.

Note that $\#_{t,t}$ for $J\ge t\ge 1$ is twice of the corresponding number in
the original level-$t$ instance. Remember $\#^{begin}_s=2^{q(s-1)}N$,
$\#^{end}_s=2\cdot 2^{q(s-2)}N$, for $J\ge s\ge 1$, and
$\#^{begin}_0=\frac{1}{J!}2^{-q}N$.

By Claim 1, $\#_{t,s}\ge \#_{t-1,s}$ for $s>t$ and by Claim 3,
$$\#_{t,t}=2\#_{t-1,t}\ge\#_{t-1,t}+\#^{end}_t=\#_{t-1,t}
+2\#^{begin}_{t-1}\ge\#_{t-1,t}+\#_{t-1,t-1},$$

\noindent it follows that $\#_t\ge\#_{t-1}$. Hence $\#_t\ge\#_{s}$, for any
$1\le s< t$.

By Claim 2, $\#_{t,s}\ge \#_s^*$ for $s>t$ and by Claim 3,
$$\#_{t,t}\ge \#_t^*+\#^{end}_{t}= \#_t^*+2\#^{begin}_{t-1}\ge
\#_t^*+\sum_{s=0}^{t-1}\#^{begin}_{s}\ge \#_t^*+\sum_{s=0}^{t-1}{\#^*_{s}},$$

\noindent it follows that $\#_t\ge\#^*$.

We conclude the algorithm could select all tests in $\mathcal T'_{t,1}$ in
their natural order, for $J\ge t\ge 1$, and select all tests in $\mathcal
T'_{0}$, finally return $\mathcal T'$.

Remember $M^*=J!2^q$, $N=J!2^{q(J+1)}$, $m^{*}=M^{*}$, and $n=(J+1)N$. In the
condition $J$ is fixed, the size of returned solution is
\begin{eqnarray*}
&|\mathcal T'|&\ge(M^*-1)(\ln N-\ln M^*)+\frac{qM^*}{8}H_{J}\\
&&= m^*(1+\frac{1}{J+1}(\frac{H_{J}}{8\ln 2}-1)-o(1))\ln n.
\end{eqnarray*}
\section{Discussion}
The author notes this is the first time to distinguish precisely the
worst case performance guarantees of two types of "greedy
algorithms" implemented by set cover criterion and by information
criterion. In fact, the author definitely shows the pattern of
instances on which ICH performs better than SGA.

In a preceding paper\cite{cl}, we proved the performance guarantee
of SGA can be $(1.5+o(1))\ln n$, and the proof can be extended to
weighted case, where each test is assigned a positive weight, and
the objective is modified as to find a test set with minimum total
weight.

In the minimum cost probe set problem\cite{bc} of bioinformatics,
tests are replaced with partitions of items. The objective is to
find a set of partitions with smallest cardinality to differentiate
all item pairs. It is easily observed that the improved
performance guarantee in this paper is still applicable to this generalized case.\\

\noindent{\bf Acknowledgements.} The author would like to thank Tao
Jiang and Tian Liu for their helpful comments.


\end{document}